\documentclass[Physsubmission, Phys]{SciPost}

\hypersetup{
    colorlinks,
    linkcolor={red!50!black},
    citecolor={blue!50!black},
    urlcolor={blue!80!black}
}

\usepackage[bitstream-charter]{mathdesign}
\DeclareSymbolFont{usualmathcal}{OMS}{cmsy}{m}{n}
\DeclareSymbolFontAlphabet{\mathcal}{usualmathcal}
\usepackage{subcaption}
\usepackage{graphicx}
\usepackage{soul}
\begin{document}

\begin{center}{\Large \textbf{
Strangeness production in small-collision systems with ALICE\\
}}\end{center}

\begin{center}
Anju Bhasin \textsuperscript{1} and Meenakshi Sharma \textsuperscript{1},
\st{} (for the ALICE Collaboration)
\end{center}

\begin{center}
{\bf 1} University of Jammu
\\

* anju.bhasin@cern.ch \\
** meenakshi.sharma@cern.ch
\end{center}

\begin{center}
\today
\end{center}


\definecolor{palegray}{gray}{0.95}
\begin{center}
\colorbox{palegray}{
  \begin{tabular}{rr}
  \begin{minipage}{0.1\textwidth}
    \includegraphics[width=23mm]{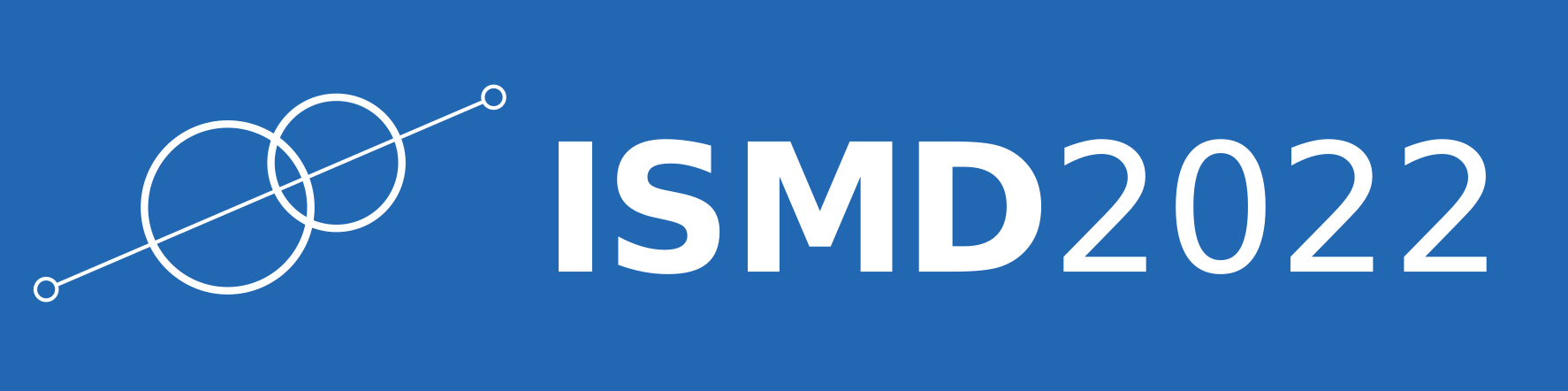}
  \end{minipage}
  &
  \begin{minipage}{0.8\textwidth}
    \begin{center}
    {\it 51st International Symposium on Multiparticle Dynamics (ISMD2022)}\\ 
    {\it Pitlochry, Scottish Highlands, 1-5 August 2022} \\
    \doi{10.21468/SciPostPhysProc.?}\\
    \end{center}
  \end{minipage}
\end{tabular}
}
\end{center}

\section*{Abstract}
{\bf
\par The main goal of the ALICE experiment is to study the physics of strongly-interacting matter, focusing on the properties of the quark-gluon plasma (QGP). The relative production of strange hadrons with respect to non-strange hadrons in heavy-ion collisions was historically considered as one of the signatures of QGP formation. However, the latest results in proton-proton (pp) and proton-lead (p--Pb) collisions have revealed an increasing trend in the yield ratio of strange hadrons to pions with the charged-particle multiplicity in the event, showing a smooth evolution across different collision systems and energies. 
\par We present the new studies which are performed with the aim of better understanding the production mechanisms for strange particles, and hence the strangeness enhancement phenomenon, in small-collision systems. In one of the recent studies, the very forward energy transported by beam remnants (spectators) and detected by the Zero Degree Calorimeters (ZDC) is used to classify events. The contribution of the effective energy and the particle multiplicity on strangeness production is studied using a multi-differential approach in order to disentangle initial and final state effects. In the second study, the origin of strangeness enhancement with multiplicity in pp has been further investigated by separating the contribution of soft and hard processes, such as jets, to strange hadron production. Techniques involving full jet reconstruction or two-particle correlations have been exploited.
The results indicate that the increased relative strangeness production emerges from the growth of the underlying event and suggest that soft (transverse to leading) processes are the dominant contribution to strange hadron production and strangeness enhancement. Further it is also seen in pp collisions that strangeness production increases with midrapidity multiplicity and there exists an anti-correlation with the effective energy.}


\section{Introduction}
\label{sec:intro}
Strangeness production is considered one of the important tools in the search and discovery of the primordial state of matter, which existed a few microseconds after the Big Bang \cite{c1,ca1,ca2}. Rafelski and Muller \cite{c1} reported for the first time that the enhancement of the relative strangeness production could be one of the signatures of a phase transition from hadronic matter to the new phase consisting of almost free quarks and gluons (QGP). Strangeness enhancement was observed for the first time at the SPS \cite{c2}, then at RHIC \cite{c3} and later at the LHC \cite{c4} at increasing collision energies.
\begin{figure}
\hspace{0.8cm}
	\begin{subfigure}[b]{0.31\textwidth}
		\includegraphics[scale=0.25]{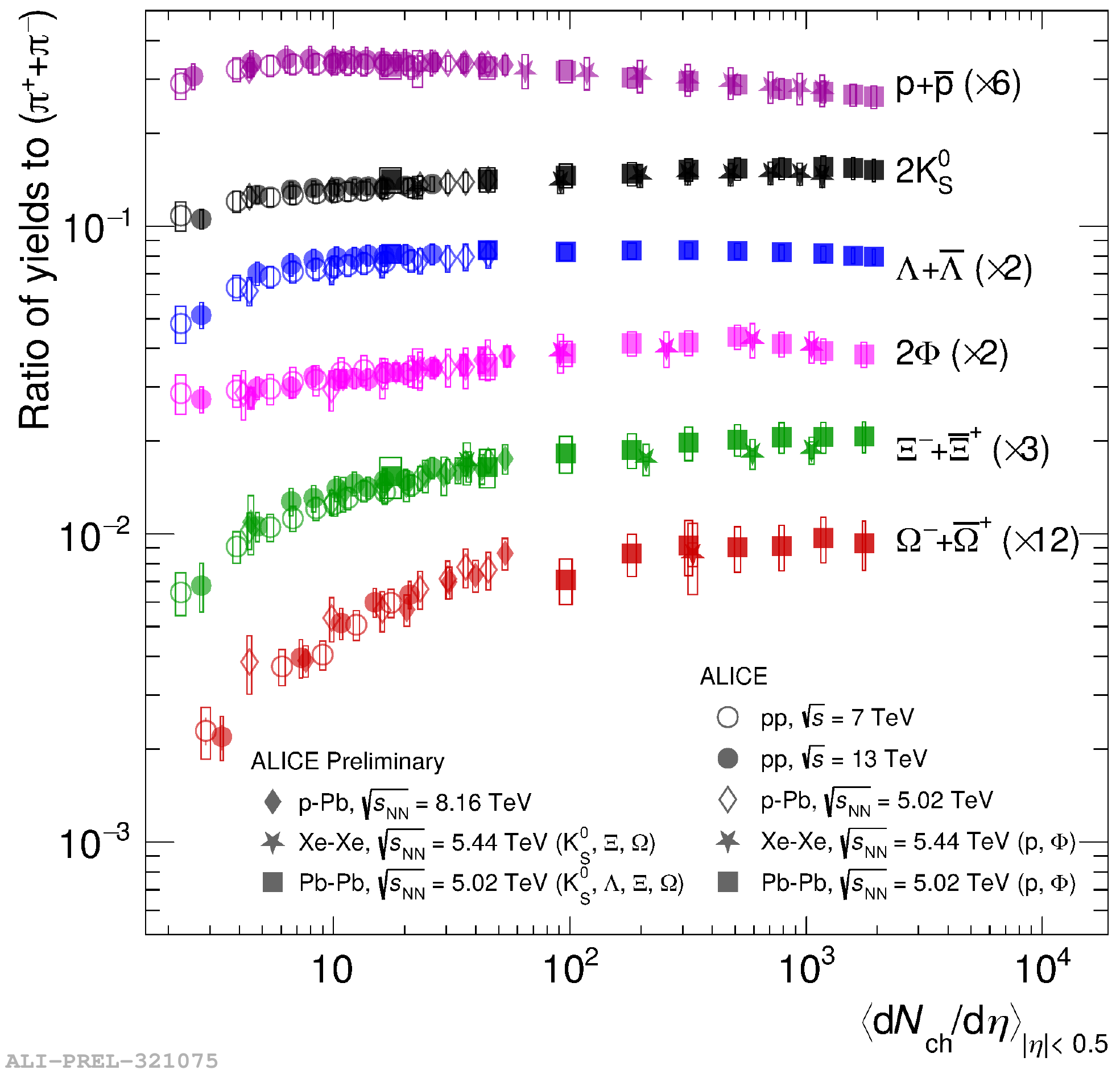}
		\caption{} \label{fig:1a}
	\end{subfigure}%
	\hspace{0.85cm}
	\begin{subfigure}[b]{0.31\textwidth}
		\includegraphics[scale=0.12]{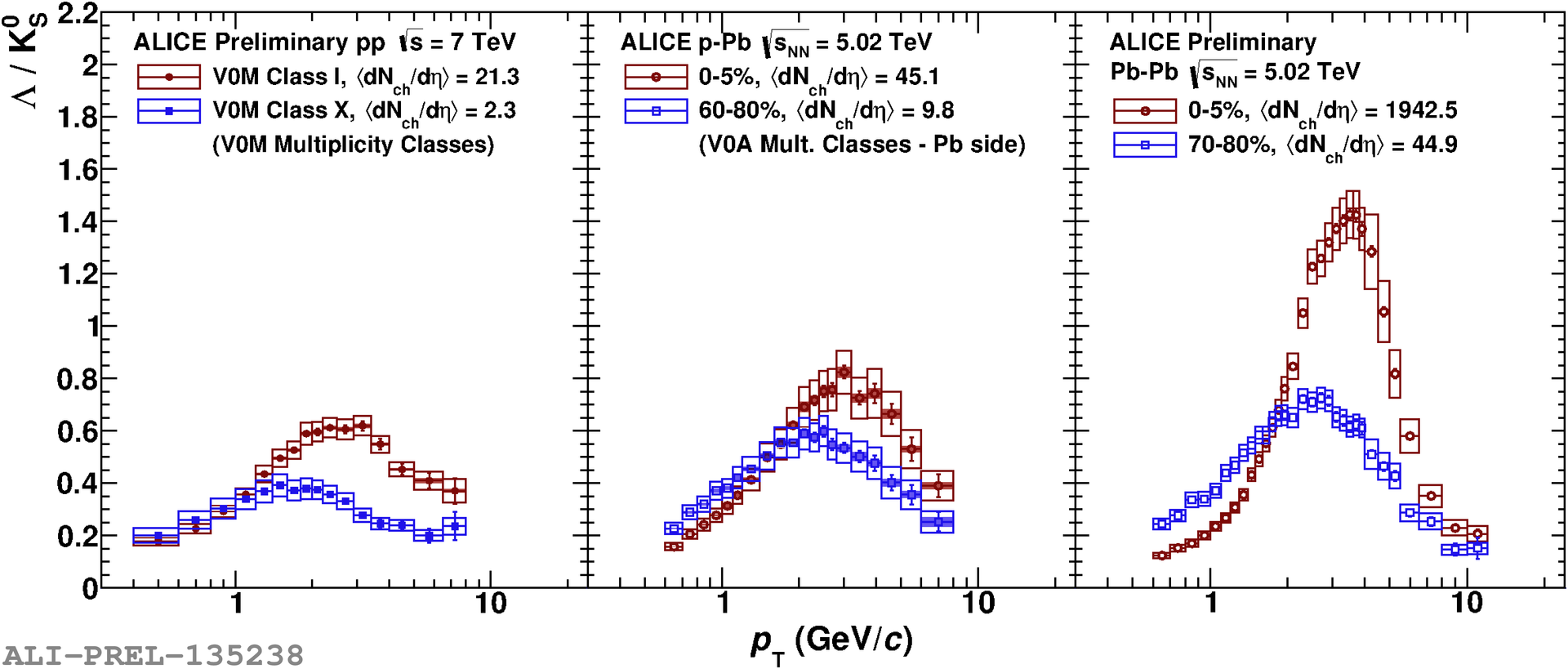}
		\caption{} \label{fig:1b}
	\end{subfigure}%
	\caption{(a) Ratio of hadron yields to pion yields in different collision systems at different collision energies as a function of $\langle \mathrm{d}N_\mathrm{ch}/\mathrm{d}\eta \rangle_{\vert \eta \vert < 0.5}$ (b) Enhancement of baryon over meson ratio at intermediate $p_{\mathrm{T}}$ across different collision system } \label{fig:1}
\end{figure}
\par The ALICE experiment has also studied strangeness production in different collision systems (pp, p--Pb and Pb--Pb) at different collision energies \cite{c7,c8,c9,c10}. The result in  figure~\ref{fig:1a} shows that the ratio of strange to non-strange hadron yields increases with charged-particle multiplicity, showing a smooth evolution across different collision system and energies.
\par It is also observed that the enhancement is larger for the particles with larger strangeness content. Several features that were observed in large collision systems and explained as due to the formation of the QGP or collective phenomenon are also observed in the small systems \cite{c12}. This includes the enhancement of $\Lambda/\mathrm{K^{0}_{s}}$ at intermediate $p_{\mathrm{T}}$ shown in  figure~\ref{fig:1b} and the evolution of the particle spectra with multiplicity and hardening of the spectra towards higher multiplicity have also been observed. These unexpected results have motivated to perform studies using ALICE detector for understanding the mechanisms responsible for such behaviour in small systems.\\
\section{Results}
\par In ALICE \cite{c5}, strange hadrons $\mathrm{K^{0}_{s}}$, $\Lambda$, $\Xi$ and $\Omega$ are reconstructed using their weak decay daughter tracks in the central pseudorapidity region \cite{cm1}. The subdetectors involved in the analyses presented here include the Inner Tracking System (ITS) and the Time Projection Chamber (TPC), which are used for particle identification (PID) and tracking. Two forward detectors placed on both sides of the ALICE interaction point, the V0A and V0C, and Zero Degree Calorimeter (ZDC) are used to classify events in multiplicity and energy event classes, respectively. A detailed description of the ALICE detector can be found in \cite{c5}. One of the three analyses presented here is perfomed in p--Pb collisions while the other two analyses exploit the multi-differential approaches to explore strangeness enhancement in pp.
\par
\begin{figure}
	\centering
	\begin{subfigure}[b]{0.31\textwidth}
		\includegraphics[scale=0.29]{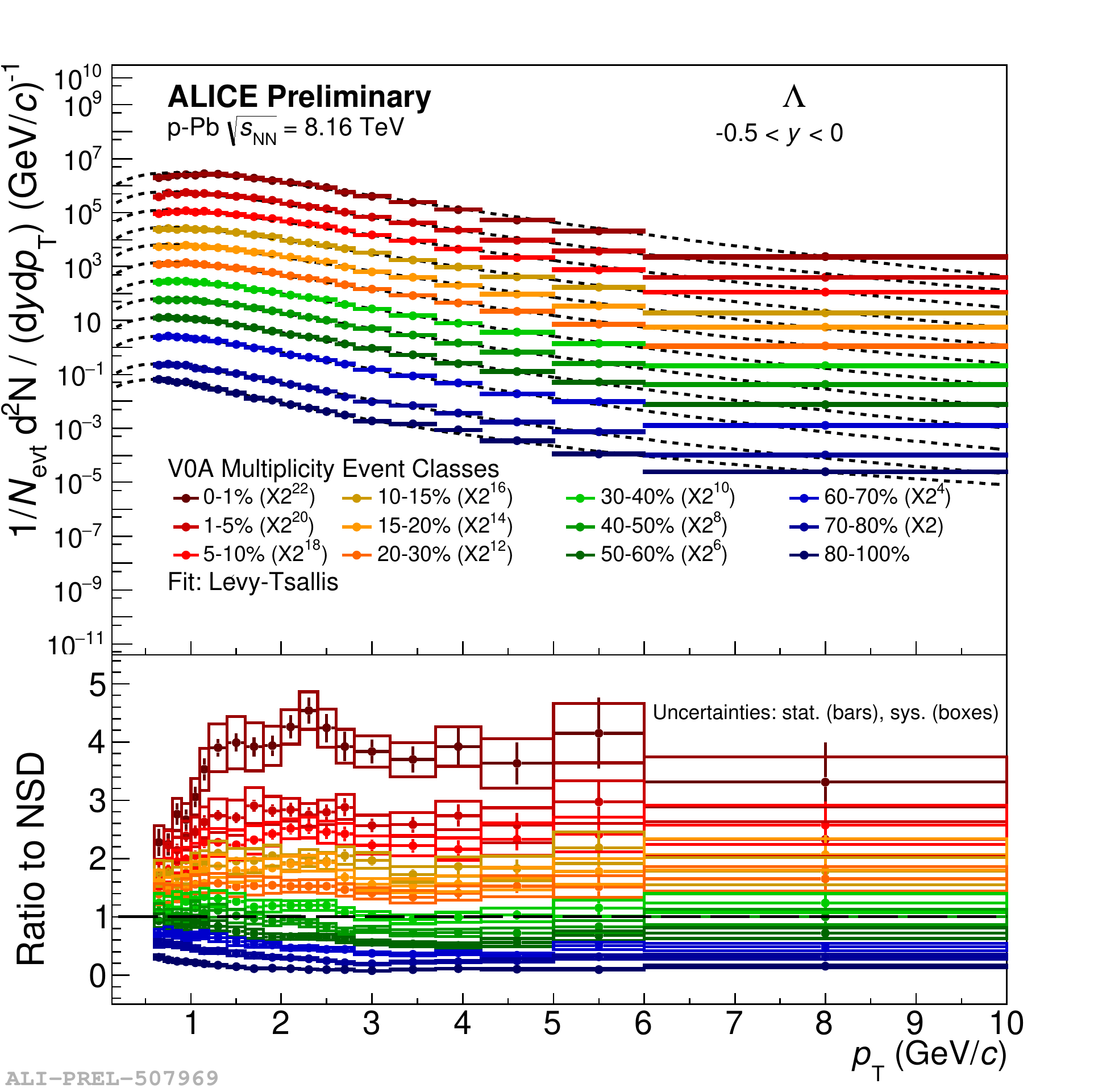}
		\caption{} \label{fig:2a}
	\end{subfigure}%
     \hspace{1.4cm}
	\begin{subfigure}[b]{0.31\textwidth}
		\includegraphics[scale=0.297]{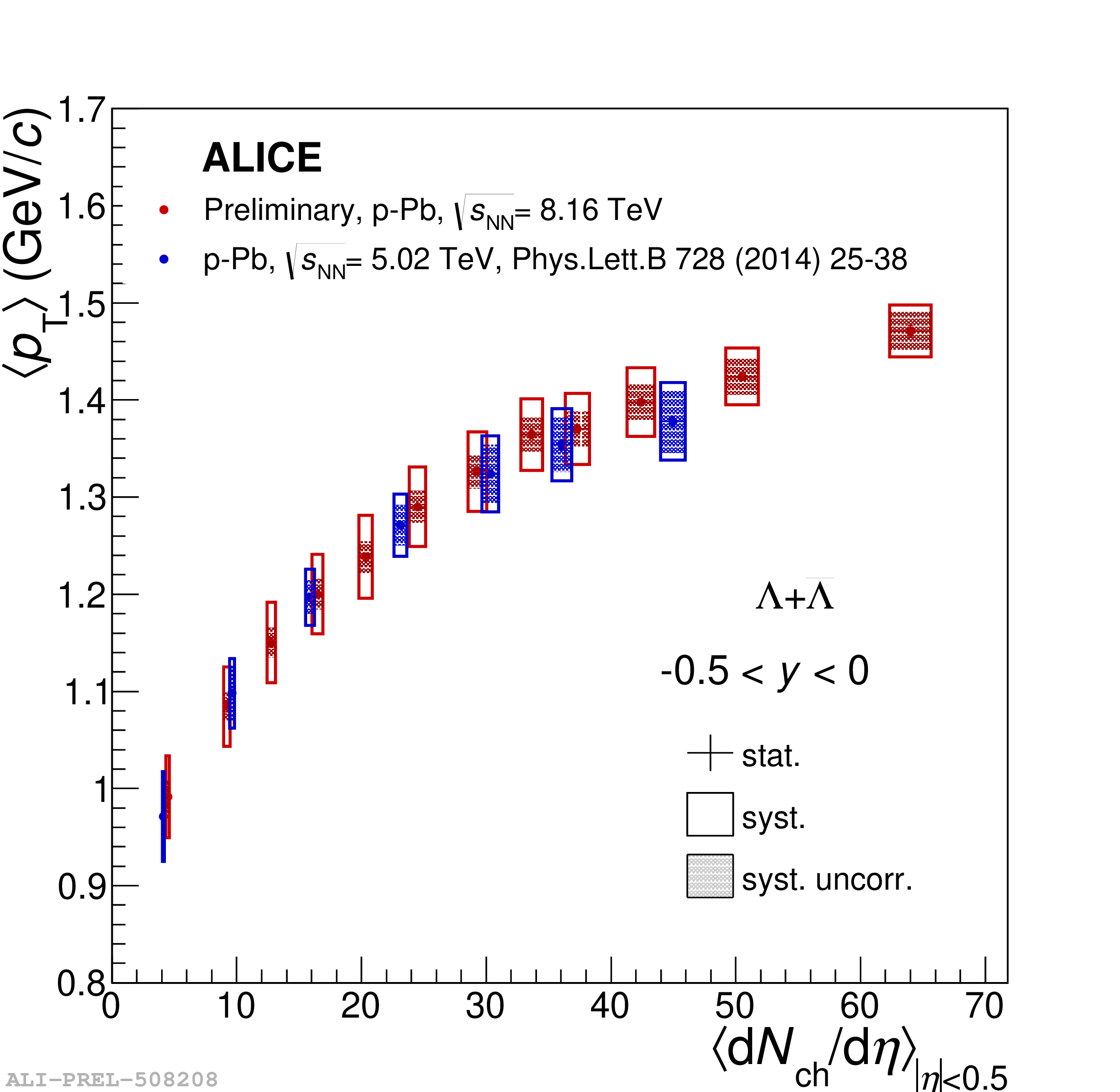}
			\caption{} \label{fig:2b}
	\end{subfigure}%

\begin{subfigure}[b]{0.31\textwidth}
	\includegraphics[scale=0.29]{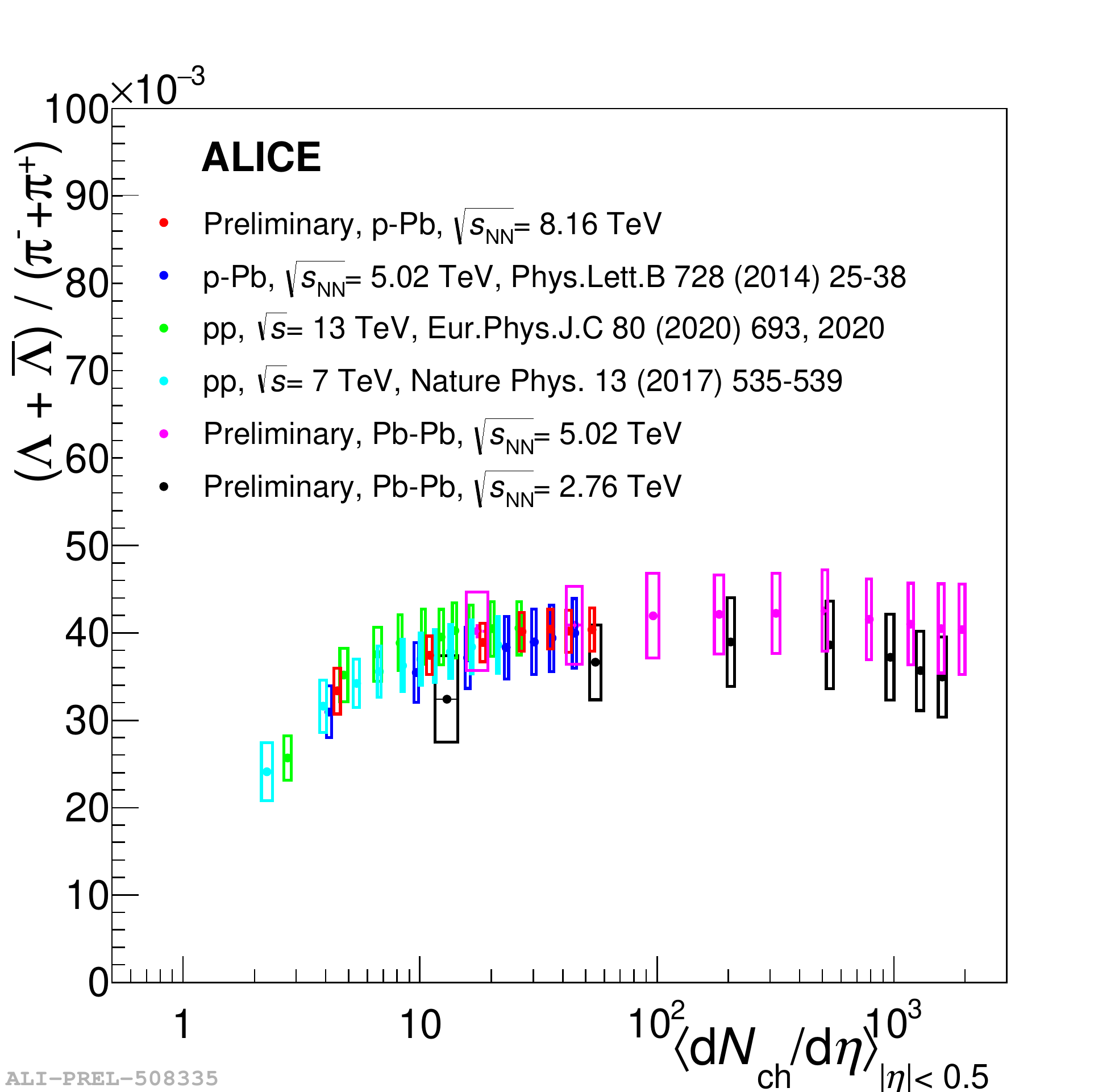}
	\caption{} \label{fig:2c}
\end{subfigure}%
\hspace{1.4cm}
\begin{subfigure}[b]{0.31\textwidth}
	\includegraphics[scale=0.29]{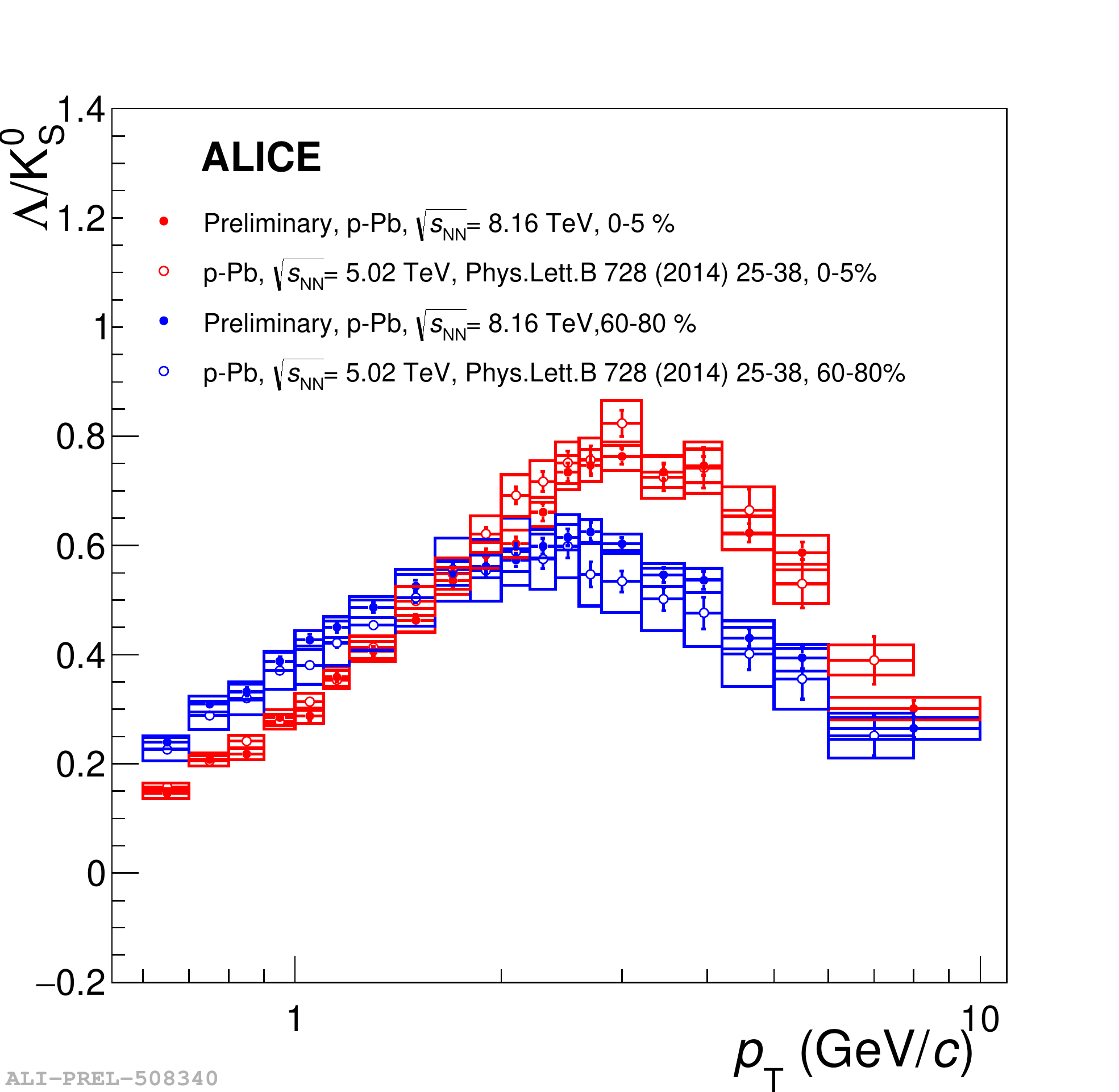}
	\caption{} \label{fig:2d}
\end{subfigure}%
\caption{(a)(Top panel): $p_{\mathrm T}$ spectra of $\Lambda$ in different V0A multiplicity classes. (Bottom panel): Ratio of $p_{\mathrm T}$ spectra in different multiplicity classes to the minimum bias $p_{\mathrm T}$ spectrum. (b) $\langle p_{T} \rangle$ of $(\Lambda+\bar{\Lambda})$ as a function of $\langle \mathrm{d}N_\mathrm{ch}/\mathrm{d}\eta \rangle_{\vert \eta \vert < 0.5}$ (c) $(\Lambda+\bar{\Lambda}) /(\pi^{+}+\pi^{-}$) as a function of $\langle \mathrm{d}N_\mathrm{ch}/\mathrm{d}\eta \rangle_{\vert \eta \vert < 0.5}$. (d) $\Lambda/\mathrm{K^{0}_{S}}$ as a function of $p_{\mathrm T}$ in 0--5$\%$ and 60--80$\%$. } \label{fig:2}
\end{figure}

 In the first analysis the measurements of $\mathrm{K^{0}_{S}}$ ($\Lambda$) are performed in the rapidity range $-0.5<y<0$ over the $p_{\mathrm T}$ range 0.2 (0.6) $<$ $p_{\mathrm T}$ $<$ 10 GeV/$c$. The yield per unit of rapidity, d\textit{N}/d\textit{y}, is obtained by integrating the measured $p_{\mathrm T}$ spectrum and estimating the yield in the unmeasured region using a Levy-Tsallis fit function.  Figure~\ref{fig:2a} shows the $p_{\mathrm T}$ spectra for $\Lambda$ in p--Pb collisions at $\sqrt{s_{\mathrm{NN}}}$= 8.16 TeV, in different V0A multiplicity classes along with their ratios to the corresponding spectrum in minimum bias collisions. The ratios suggest that the  $p_{\mathrm T}$ spectra get harder with increasing multiplicity. This arguement is more clear from  figure~\ref{fig:2b}, where the average transverse momentum increase as a function of the average charged particle pseudorapidity density at midrapidity $\langle \mathrm{d}N_\mathrm{ch}/\mathrm{d}\eta \rangle_{\vert \eta \vert < 0.5}$. The $p_{\mathrm T}$ integrated yield ratio $(\Lambda+\bar{\Lambda})/(\pi^{+}+\pi^{-}$) as a function of $\langle \mathrm{d}N_\mathrm{ch}/\mathrm{d}\eta \rangle_{\vert \eta \vert < 0.5}$ in different systems is shown in figure~\ref{fig:2c}. A smooth evolution is observed for this ratio going from from low to high multiplicity. Moreover, the baryon over meson ratio $\Lambda/\mathrm{K^{0}_{S}}$ in two different multiplicity intervals and energies is shown in figure~\ref{fig:2d}. A significant peak is observed at intermediate $p_{\mathrm T}$, which can be explained by effects of radial flow combined with processes like recombination during the hadronisation of the created QGP. The measured trend with multiplicity shows a more pronounced peak in high-multiplicity (0--5$\%$) as compared to low multiplicity (60--80$\%$) p--Pb collisions and no dependence on the collision energy within uncertainties.
\begin{figure}
	\centering 
	\begin{subfigure}[b]{0.251\textwidth}
		\includegraphics[scale=0.34]{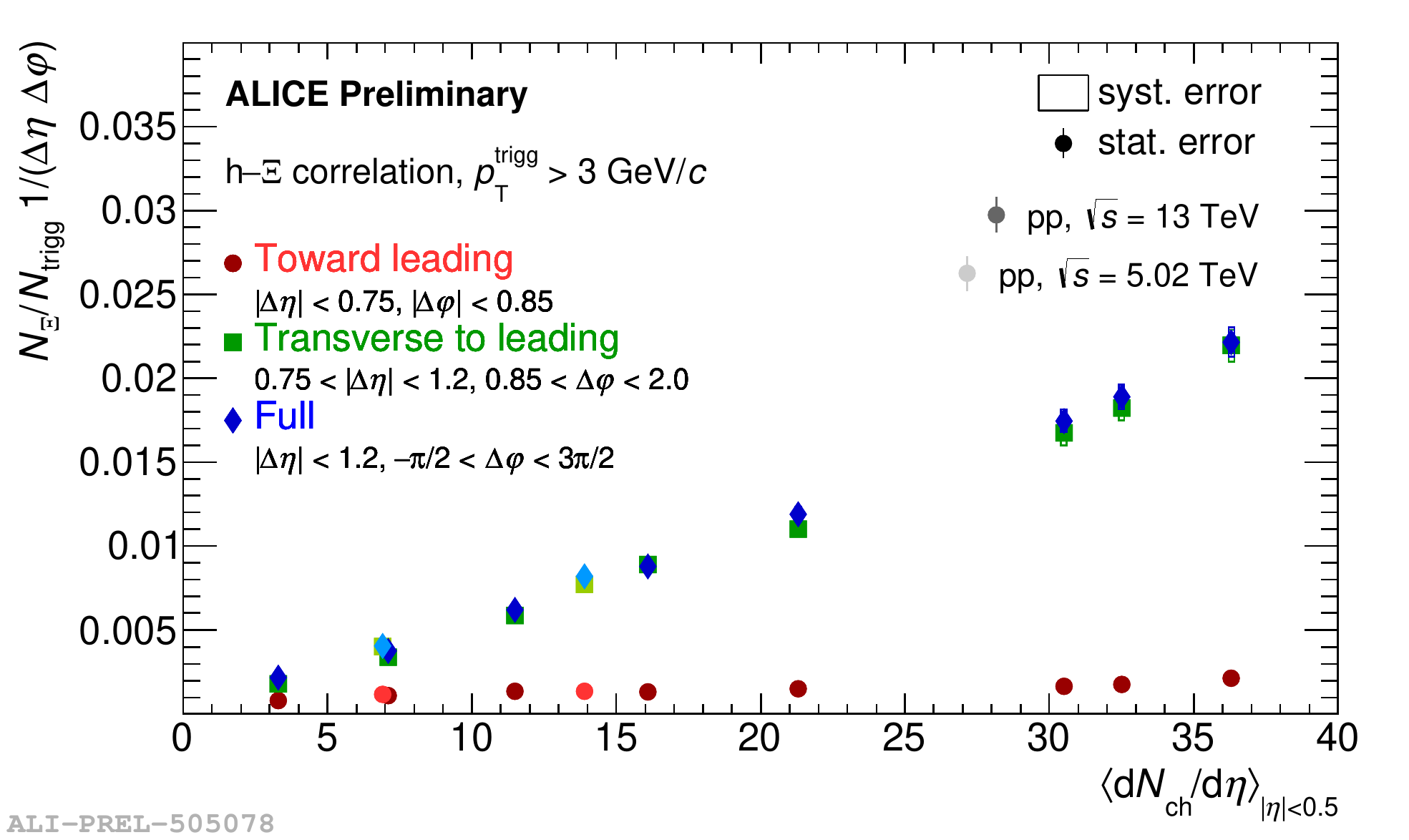}
		\caption{} \label{fig:3a}
	\end{subfigure}%
	\hspace{2.8cm}
	\begin{subfigure}[b]{0.31\textwidth}
		\includegraphics[scale=0.30]{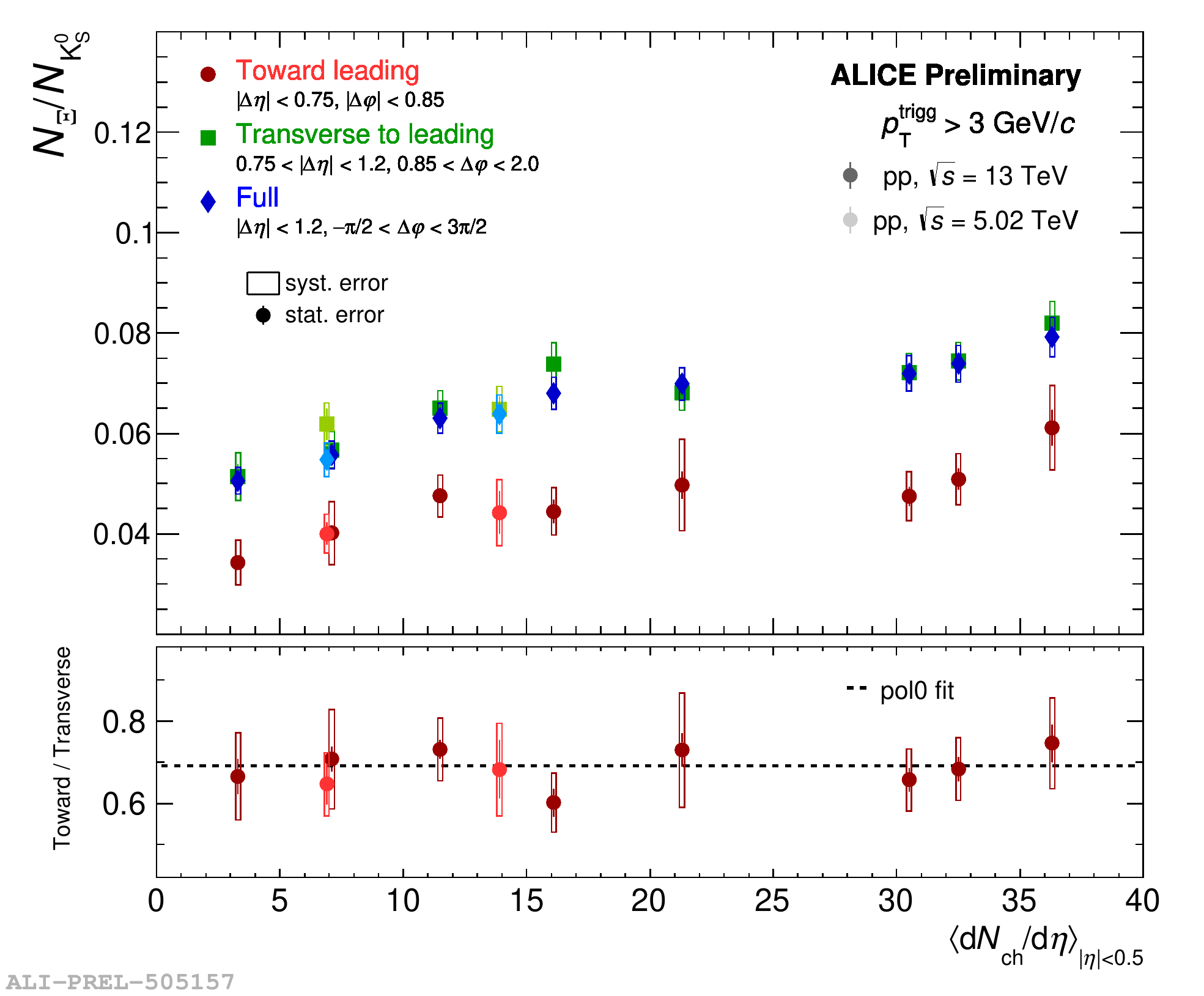}
		\caption{} \label{fig:3b}
	\end{subfigure}
	\caption{(a) $\mathrm \Xi$ yields per trigger particle and per unit of $\mathrm \Delta\eta\Delta\phi$ as a function of the charged-particle multiplicity produced at midrapidity. (b) $\mathrm \Xi$/$\mathrm{K^{0}_{s}}$ yield ratio as a function of the charged-particle multiplicity produced at midrapidity} \label{fig:3}
\end{figure}
\begin{figure}
	\centering
	\hspace{-2.8cm}
	\begin{subfigure}[b]{0.31\textwidth}
\includegraphics[scale=0.42]{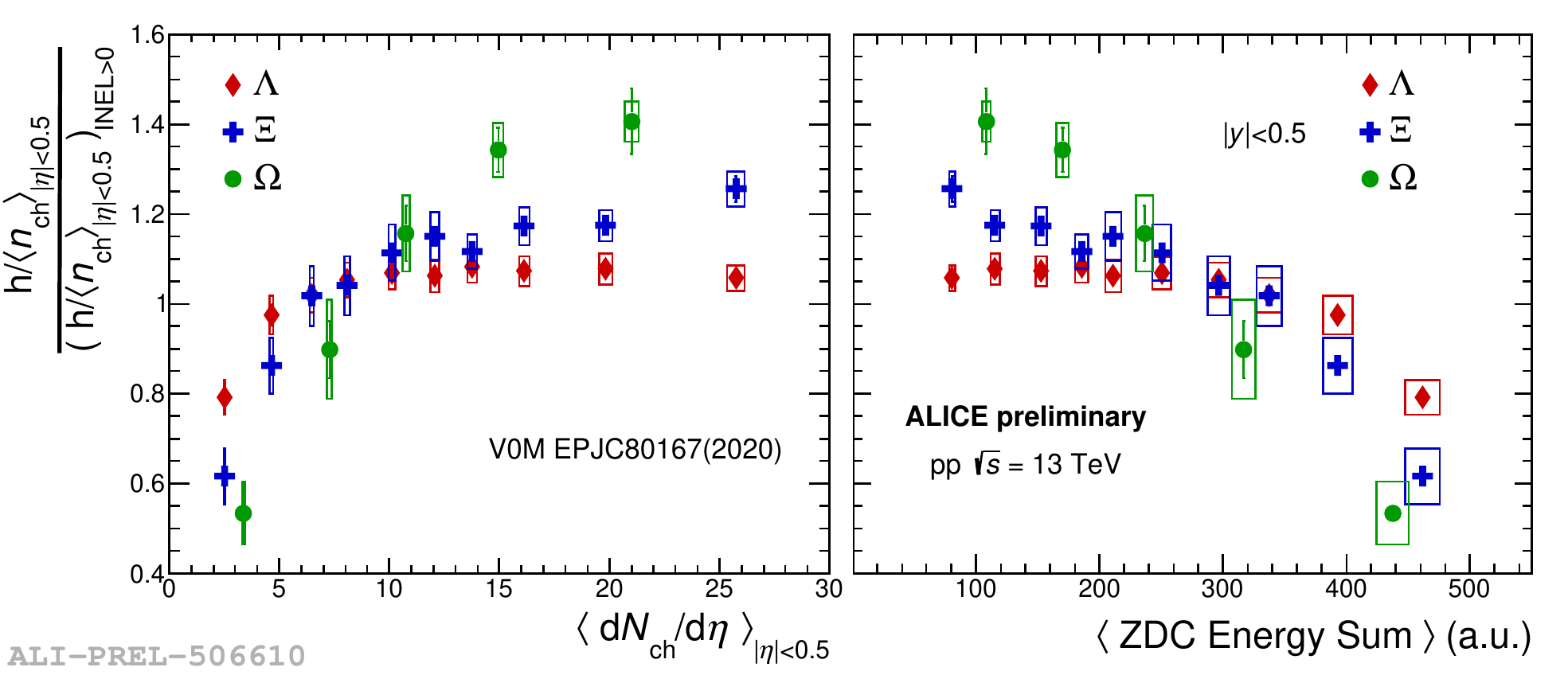}
\caption{} \label{fig:4a}
\end{subfigure}

\hspace{-2.8cm}
\begin{subfigure}[b]{0.31\textwidth}
	\includegraphics[scale=0.42]{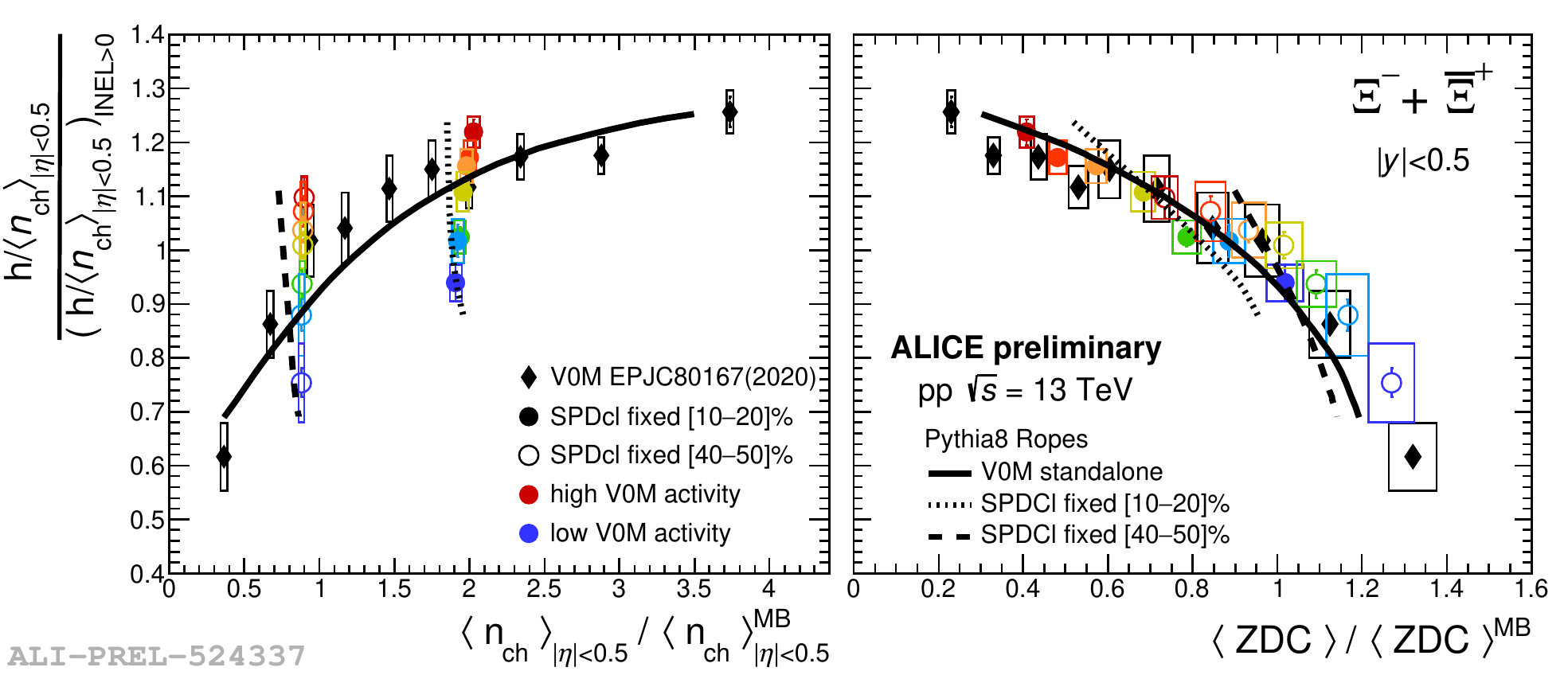}
	\caption{} \label{fig:4b}
\end{subfigure}

\hspace{-2.8cm}
\begin{subfigure}[b]{0.31\textwidth}
	\includegraphics[scale=0.42]{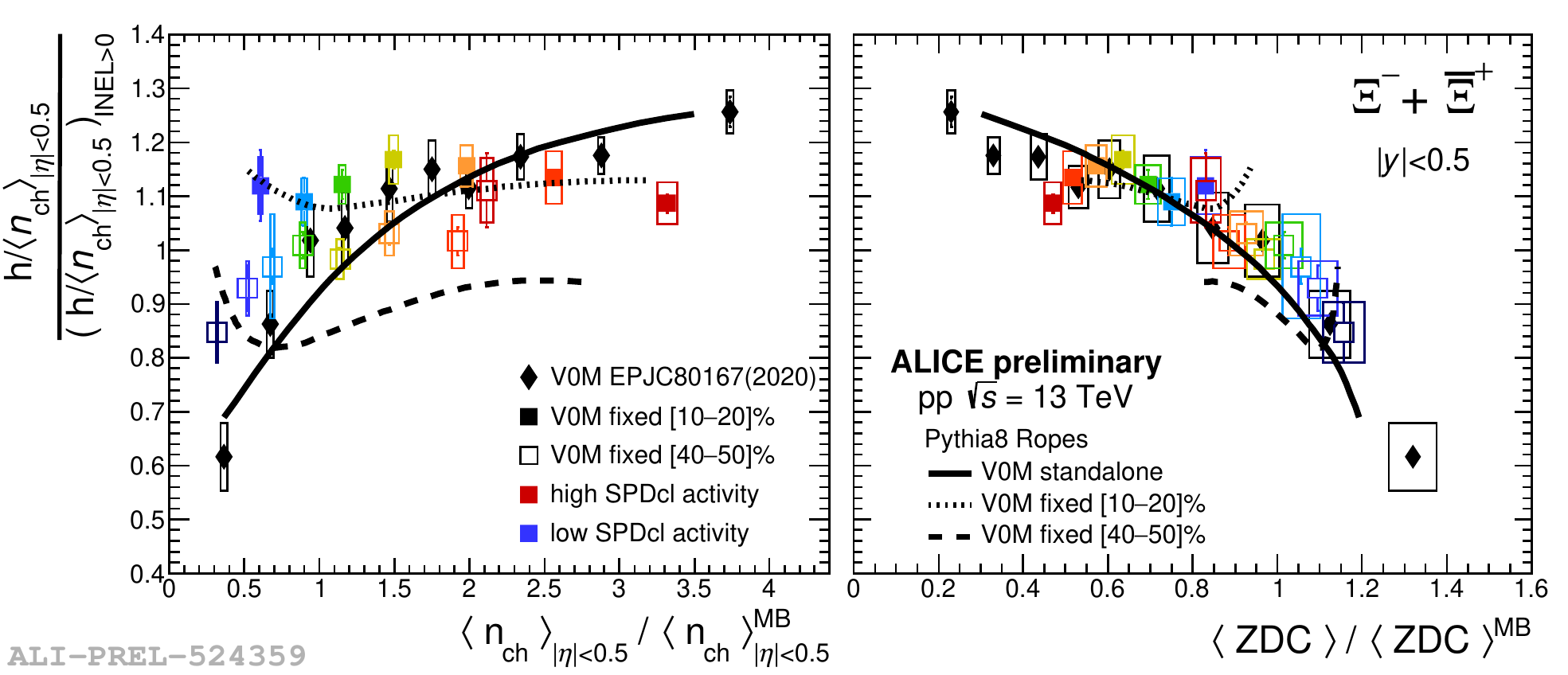}
	\caption{} \label{fig:4c}
\end{subfigure}%
\caption{(a) The yield of strange hadrons normalised to the charged particle multiplicity as a function of multiplicity at midrapidity (left plot) and ZDC sum energy sum (right plot) (b) $\Xi^{\pm}$ yield normalised to the charged particle multiplicity fixing the multiplicity at midrapidity (c)  $\Xi^{\pm}$ yield normalised to the charged particle multiplicity in events with ZDC energy deposits fixed in a small range} \label{fig:4}
\end{figure}
\par The second analysis involves the angular correlation method to separate $\mathrm{K^{0}_{s}}$ and $\Xi$ hadrons produced in toward leading (hard processes) from those produced transverse to leading (soft processes), where the hard processes are characterised by large momentum transfer. The particles produced in the near-side jet region are characterised by a small angular separation from the leading particle of the jet, which is identified as the particle with the highest transverse momentum in the collision with $p_{\mathrm{T}}$ $>$  3 GeV/\textit{c}. The $\Xi$ yields per trigger particle and per unit of $\Delta\eta\Delta\phi$ (where $\Delta\eta$ = $\eta_{Trigger} - \eta_{Associated} $ and $\Delta\phi$ = $\phi_{Trigger} - \phi_{Associated} $) are displayed in figure~\ref{fig:3a} as a function of the charged-particle multiplicity produced at midrapidity. The full and transverse to leading yields increase with multiplicity, while the toward leading yields show a very mild to no dependence on particle production at midrapidity. Similar results are also obtained for $\mathrm{K^{0}_{s}}$ yields. In figure~\ref{fig:3b}, $\Xi$ /$\mathrm{K^{0}_{s}}$ yield ratio as a function of charged-particle multiplicity produced at midrapidity is shown. These results suggest that transverse to leading (soft) processes are the dominant contribution to strange particle production and strangeness enhancement is also observed in pp collisions.
\par The fraction of the initial energy spent in the hadronization process is known as effective energy. Due to the leading baryon effect, there is a high probability to emit baryons in the forward direction with high longitudinal momentum, carrying away a considerable fraction of the total available energy. Therefore, the effective energy is always less than the initial collision energy \cite{c6}. In ALICE, the Zero Degree Calorimeters (ZDCs) are used to reconstruct the energy of the leading nucleons and define the effective energy classes using the relation, $\mathrm E_{eff} = \sqrt{s} - \mathrm E_{|\eta|>8}$. The events are classified in effective energy and multiplicity percentile classes, using the ZDC  and V0 detector, respectively. Both Monte-Carlo simulation and data have confirmed that the effective energy and multiplicity are correlated.

\par The self-normalized ratio of yields to the average charged-particle multiplicity (in INEL$>$0) with multiplicity selected through both V0M and ZDC are shown in Figure~\ref{fig:4a}. The figure clearly shows that the ratio of strange hadrons yield to the charged particle multiplicity increases with multiplicity and is anti-correlated with very forwarded energy.
To distinguish the initial and final state contributions to strange particle production, V0M and ZDC combined classes are exploited using two approaches. From figure \ref{fig:4b} and figure \ref{fig:4c}, strangeness enhancement in pp collsions is observed at fixed final state multiplicity at midrapidity and shows a strong correlation with the effective energy, which reflects the initial state of the collision.
\section{Conclusions}
\label{sec:another}
\par The ALICE collaboration has performed studies for understanding strange particle production in high-multiplicity pp and p--Pb collisions. The measurements performed for different collision systems are consistent with each other at similar multiplicities and show a common increasing trend. This suggests  that the particle production is independent of collision system and collision energy and mainly driven by $\langle \mathrm{d}N_\mathrm{ch}/\mathrm{d}\eta \rangle_{\vert \eta \vert < 0.5}$. 
\par ALICE exploits multi-differential approach to disentangle the contribution of multiplicity and effective energy to strange particles production. It has been observed that there is an increase in strangeness production at fixed midrapidity multiplicity. Moreover, the strangeness production shows a strong correlation with the effective energy following a universal trend with the leading energy detected by the ZDC. Additionally, the study of strange hadrons toward leading and transverse to leading suggests that soft processes are the dominant contribution to strange particle production. 
\section*{Acknowledgements}
We thank and acknowledge the financial support by the Department of Science and Technology (DST), Government of India. 


\newpage





\end{document}